\documentstyle[12pt]{article}
\advance \topmargin by -\headheight
\advance \topmargin by -\headsep
\evensidemargin \oddsidemargin
\begin{document}
\begin{center}
{\bf Gravity, Cosmology and Particle Physics 
\\
without the Cosmological Constant Problem}\footnote{This essay received an 
"honorable mention" from the Gravity Research Foundation, 1998.}
 \end{center}

\bigskip
\begin{center}

\bigskip

E.I.Guendelman\footnote{Electronic address: GUENDEL@BGUmail.BGU.AC.IL} and
        A.B.Kaganovich\footnote{Electronic address: ALEXK@BGUmail.BGU.AC.IL}

{\it Physics Department, Ben Gurion University, Beer
Sheva, Israel}
\end{center}
\bigskip

\begin{abstract}

This essay elucidates  recent achievements of the "nongravitating vacuum 
energy" (NGVE) theory" which has the feature that a shift of the 
Lagrangian density by a constant does not affect dynamics. In the first 
order formalism, a constraint appears that enforces the vanishing of the 
cosmological constant $\Lambda$. Standard dynamics of gauge unified 
theories (including fermions) and their SSB appear if a four index field 
strength condensate is present. At a vacuum state there is exact balance 
to zero of the gauge fields condensate and the original scalar fields 
potential. As a result it is possible to combine the solution of the 
$\Lambda$ problem with inflation and transition to a $\Lambda =0$ phase 
without fine tuning after a reheating period. The model opens new 
possibilities for a solution of the hierarchy problem.

 \end{abstract}

\pagebreak

A familiar notion in theoretical physics is that the "origin of energy is
not important". Gravity, if formulated in the standard way opposes such
idea: in General Relativity (GR) the action $\int L\sqrt{-g}d^{4}x$ is not 
invariant under the shift
$L\rightarrow L+const$, which in fact gives rise to a cosmological 
constant in the equations of motion.
Here we want to restore $L\rightarrow L+const$ as a symmetry, even if 
gravity is introduced.
For this we change $\sqrt{-g}d^{4}x$ by the volume form 
$d\varphi_{1}\wedge
d\varphi_{2}\wedge
d\varphi_{3}\wedge
d\varphi_{4}\equiv\frac{\Phi}{4!}d^{4}x$ where
$\Phi \equiv \varepsilon_{abcd}
\varepsilon^{\mu\nu\alpha\beta}
(\partial_{\mu}\varphi_{a})
(\partial_{\nu}\varphi_{b})
(\partial_{\alpha}\varphi_{c})
(\partial_{\beta}\varphi_{d})$, and $\varphi_{a}\quad (a=1,...,4)$, 
the measure fields, are independent degrees of freedom. Since 
$\Phi$ is a total
divergence, the symmetry $L\rightarrow L+const$, i.e. the NGVE 
principle, is automatically satisfied \cite{GK1}.

Having allowed the measure not to be necessarily $\sqrt{-g}$, but to be 
determined
dynamically, it seems natural also to allow the connection to be determined
dynamically and not to assume it to be the Christoffel symbol from the
beginning. This is the case in the first order formalism, which leads to
the resolution of the cosmological constant problem in the NGVE theory
\cite{GK2}-\cite{GK3}.

We consider then the action
$S=\int\Phi Ld^{4}x$
and assume that $L$ does not contain the measure fields , that is the
fields by means of which $\Phi$ is defined. If this condition is
satisfied then the theory has an additional symmetry \cite{GK1}. Our
choice for the total Lagrangian density is
$L=\kappa^{-1}R(\Gamma,g)+L_{m}$,
where $L_{m}$ is the matter Lagrangian density and $R(\Gamma,g)$ is the
scalar  curvature
$R(\Gamma,g)=g^{\mu\nu}R_{\mu\nu}(\Gamma)$ of the space-time of the
affine connection  $\Gamma^{\mu}_{\alpha\beta}$: \,
$R_{\mu\nu}(\Gamma)=R^{\alpha}_{\mu\nu\alpha}(\Gamma)$, \,
$R^{\lambda}_{\mu\nu\sigma}(\Gamma)\equiv 
\Gamma^{\lambda}_{\mu\nu,\sigma}-\Gamma^{\lambda}_{\mu\sigma,\nu}+
\Gamma^{\lambda}_{\alpha\sigma}\Gamma^{\alpha}_{\mu\nu}-
\Gamma^{\lambda}_{\alpha\nu}\Gamma^{\alpha}_{\mu\sigma}$.
This curvature tensor is invariant under the
$\lambda$- transformation \cite{A}
$\Gamma^{\prime 
\mu}_{\alpha\beta}=\Gamma^{\mu}_{\alpha\beta}+
\delta^{\mu}_{\alpha}\lambda,_{\beta}$. \,
In the
NGVE-theory, it  allows us to eliminate the contribution to the
torsion which appears as a result of introduction of the new measure.
However, even after this
still 
there is the non metric contribution to the 
connection related to the measure (it is expressed in terms of 
derivatives the scalar field $\chi\equiv\Phi/\sqrt{-g}$).

In addition to this, in the vacuum and in some matter models, the theory
possesses a local symmetry which plays a major role. This symmetry
consists of a conformal transformation of the metric
$g_{\mu\nu}(x)=J^{-1}(x)g^{\prime}_{\mu\nu}(x)$ accompanied by a 
corresponding
diffeomorphism $\varphi_{a}\longrightarrow\varphi^{\prime}_{a}=
\varphi^{\prime}_{a}(\varphi_{b})$ in the space of the scalar fields
$\varphi_{a}$ such that  $J=
Det(\frac{\partial\varphi^{\prime}_{a}}{\partial\varphi_{b}})$.
Then for $\Phi$ we have: $ \Phi(x)=J^{-1}(x)\Phi^{\prime}(x)$. In the
presence of fermions this symmetry is appropriately generalized \cite{GK2}.
For models where it
holds, it is possible to choose the gauge where the measure $\Phi$
coincides with $\sqrt{-g}$, the measure of GR. This is
why we call this symmetry {\em "local Einstein symmetry"} (LES).

 Varying the action with respect to $\varphi_{a}$ we get
$A^{\mu}_{b}\partial_{\mu}\lbrack -\frac{1}{\kappa}R(\Gamma,g)+
L_{m}\rbrack =0$ where $A^{\mu}_{b}=\varepsilon_{acdb}
\varepsilon^{\alpha\beta\gamma\mu}
(\partial_{\alpha}\varphi_{a})
(\partial_{\beta}\varphi_{c})
(\partial_{\gamma}\varphi_{d})$.
If $Det (A_{b}^{\mu}) =
\frac{4^{-4}}{4!}\Phi^{3}\neq 0$ then
\begin{equation}
-\frac{1}{\kappa}R(\Gamma,g)+L_{m}=M=const
\label{1}
\end{equation}

Performing the variation with respect to
$g^{\mu\nu}$ we get (here for simplicity we don't consider fermions)
 \begin{equation}
-\frac{1}{\kappa}R_{\mu\nu}(\Gamma)+\frac{\partial L}{\partial g^{\mu\nu}}=0
\label{2}
\end{equation}

Contracting eq.(\ref{2}) with $g^{\mu\nu}$ and making use eq.(\ref{1}) we
get the constraint
\begin{equation}
g^{\mu\nu}\frac{\partial(L_{m}-M)}{\partial g^{\mu\nu}}-(L_{m}-M)=0
\label{3}
\end{equation}

For the cases where the LES is an exact symmetry, we
can eliminate the mentioned above $\chi$-contribution to the connection.
Indeed, for  $J=\chi$ we get $\chi^{\prime}\equiv 1$ and
$\Gamma^{\prime \alpha}_{\mu\nu}=
\{ ^{\alpha}_{\mu\nu}\}^{\prime}$, where
$\{ ^{\alpha}_{\mu\nu}\}^{\prime}$
 are the Christoffel's coefficients corresponding
to the new metric $g^{\prime}_{\mu\nu}$. In this gauge the affine space-time
becomes a Riemannian space-time.

When applying the theory to the matter model of a single scalar field with a 
potential $V(\varphi)$, the constraint (\ref{3}) implies $V(\varphi)+M=0$, 
which means that $\varphi$ is a constant (the equation of motion of 
$\varphi$ implies also that $V^{\prime}(\varphi)=0$). Since 
$\varphi= constant$, Eq. (\ref{2}) implies $R_{\mu\nu}(\Gamma,g)=0$ and 
also Eq. (\ref{1}) implies $R(\Gamma,g)=0$, if $V(\phi)+M=0$ is taken 
into account. $V(\phi)+M=0$ dictates the vanishing of the LES violating 
terms and disappearance of dynamics of $\varphi$, so LES is effectively 
restored on the mass shell. Choosing the gauge $\chi =1$ we see that the 
potential does not have a gravitational effect since the standard Ricci 
tensor vanishes and flat space-time remains the only natural vacuum.

The above solution of the cosmological constant problem is at the prize 
of the elimination of a possible scalar field dynamics. We will see now 
that the introduction of a 4-index field strength can restore normal 
scalar, gauge and fermion dynamics and as a bonus provide 
\cite{GK3},\cite{GK4}: (i) {\em the possibility of inflation in the early 
universe with a transition (after reheating) to a $\Lambda =0$ phase without 
fine tuning} and
(ii) {\em a solution to the hierarchy problem in the context of unified 
gauge dynamics with SSB}.

A four index field strength is derived from a three index gauge 
potential according to 
$F_{\mu\nu\alpha\beta}=\partial_{[\mu}A_{\nu\alpha\beta]}$.
The physical scenario we have in mind is one where all gauge fields, 
including $A_{\nu\alpha\beta}$ are treated in a unified way. The possible 
physical origin of $A_{\nu\alpha\beta}$ can be for example an effective 
way to describe the condensation of a vector gauge field in some extra 
dimensions \cite{G}. In this picture for example this means that all 
gauge fields should appear in a combination having all the same 
homogeneity properties with respect to conformal transformations of the 
metric. This is achieved if all dependence on field strengths is through 
the "gauge fields complex" \,
$y=F^{a}_{\mu\nu}F^{a\mu\nu}+
\frac{\varepsilon^{\mu\nu\alpha\beta}}{\sqrt{-g}}\partial_{\mu}
A_{\nu\alpha\beta}$.

Considering for illustration only one vector gauge field $\tilde{A}_{\mu}$ 
and 
a charged scalar field $\phi$, we take a generic action satisfying the 
above requirements in the unitary gauge ($\phi =\phi^{*};\, |\phi|=
\frac{1}{\sqrt{2}}\varphi$)
\begin{equation}
S=\int\Phi d^{4}x\left[-\frac{1}{\kappa}R(\Gamma ,g)-m^{4}f(u)
+\frac{1}{2}g^{\mu\nu}\partial_{\mu}\varphi\partial_{\nu}\varphi
-V(\varphi)+\frac{1}{2}\tilde{e}^{2}\varphi^{2}g^{\mu\nu}\tilde{A}_{\mu}
\tilde{A}_{\nu}\right],
\label{5}
\end{equation} 
where $u\equiv y/m^{4}$,\, $m$ is a mass parameter 
and $f(u)$ is a nonspecified function which has to have an extremum at 
some point $u=u_{0}>0$ to provide physically reasonable consequences (see 
below).   

The equations of motion obtained from variating $A_{\nu\alpha\beta}$ imply
$\chi f^{\prime}=\omega =constant$
where $\omega$ is a dimensionless integration constant.
The constraint (\ref{3}) becomes now
\begin{equation}
-2uf^{\prime}(u)+f(u)+\frac{1}{m^{4}}[V(\varphi)+M]=0,
\label{7}
\end{equation}
which allows to find $u=u(\varphi)$.

One can then see that all equations can be put in the standard GR, scalar 
field and gauge field form if we make the conformal transformation to the 
"Einstein frame"
$\overline{g}_{\mu\nu}=\chi g_{\mu\nu}$.
In the Einstein frame, the scalar field acquires an effective potential 
\begin{equation}
V_{eff}(\varphi)=\frac{y}{\omega}\left(f^{\prime}(u)\right)^{2}
\label{9}
\end{equation}

If there is a point $u=u_{0}$ where 
$f^{\prime}(u_{0})=0$, then if $y_{0}/\omega >0$, such a state is a 
stable vacuum of the theory. This vacuum state is defined by the gauge 
and scalar condensates 
$(u_{0}, \varphi_{0})$ connected 
by the relation $f(u_{0})+\frac{1}{m^{4}}[V(\varphi_{0})+M]=0$,
representing the exact cancellation of the contributions to the vacuum 
energy of the scalar field and of gauge field condensate. Therefore the 
effective cosmological constant in this vacuum becomes zero without fine 
tuning.

One can see also that $\frac{dV_{eff}}{d\varphi}=\frac{1}{\omega}
\frac{df}{du}\frac{dV}{d\varphi}$, so another extremum where 
$V^{\prime}=0$ for example can serve as a phase with nonzero effective 
$\Lambda$ and therefore inflation becomes possible as well. This vacuum 
is smoothly connected by dynamical evolution of the scalar field, with 
$\Lambda =0$ one, thus providing a way to achieve 
inflation and transition (after standard reheating period) to $\Lambda =0$
phase without fine tuning.

Stability of gauge fields requires $\omega >0$. In this case theory 
acquires canonical form if the new fields and couplings are defined
$A_{\mu}=2\sqrt{\omega}\tilde{A}_{\mu}$, \, 
$e=\frac{\tilde{e}}{2\sqrt{\omega}}$. Appearance of the VEV of the scalar 
field $\varphi_{0}$ leads to 
the standard Higgs mechanism.

Fermions can also be introduced \cite{GK4} in such a way that normal 
massive propagation is obtained in the Einstein frame. The resulting fermion 
mass 
is proportional to $\frac{\varphi_{0}}{\sqrt{u_{0}\omega}}$. The gauge boson 
mass, depending on $e$ also goes as $\propto{1}/{\sqrt{\omega}}$. So we 
see that a big value of the integration constant $\omega$ pushes both 
effective masses and gauge coupling constants to small values, thus 
providing a new approach to the solution of the hierarchy problem. 
Furthermore we see that fermion masses include additional factor 
$u^{-1/2}_{0}$. Therefore, if the gauge complex condensate $u_{0}$ is 
big enough, it can explain why fermion masses are much less then boson 
ones. 

Finally, there is no obstacle for the construction of a realistic unified 
theories (like electroweak, QCD, GUT) along the lines of the simple example 
displayed above \cite{GK4}. The common feature of such theories is the fact 
that the 
stable vacuum developed after SSB has zero effective cosmological constant.

\end{document}